\documentclass[12pt]{article}
\usepackage[utf8x]{inputenc}
\usepackage[T1]{fontenc}
\usepackage{amssymb,amsmath,mathtools,amsfonts}
\usepackage{graphicx}
\usepackage{lmodern}
\usepackage{float}
\usepackage[colorlinks=true, allcolors=blue]{hyperref}
\usepackage{tabularx,booktabs}
\usepackage{abstract}
\usepackage{caption}
\usepackage{titlesec}
\usepackage{adjustbox}
\usepackage[backend=biber,sorting=none]{biblatex}
\evensidemargin\oddsidemargin \hoffset-22.4mm \voffset-28.4mm
\graphicspath{{./fig/}}

\title{Geometric sensitivity of modal parameters in wind instrument models: a case study on {s}axophone intonation}
\author{Nathan Szwarcberg\textsuperscript{a, b}\footnote{Corresponding author. \\
E-mail address: nathan.szwarcberg@buffetcrampon.com (N.\ Szwarcberg)}, 
Tom Colinot\textsuperscript{a}, 
Christophe Vergez\textsuperscript{b}, 
Michaël Jousserand\textsuperscript{a}\\
\small
\textsuperscript{a} Buffet Crampon, 5 Rue Maurice Berteaux, 78711 Mantes-la-Ville, France\\
\small
\textsuperscript{b} Aix Marseille Univ, CNRS, Centrale Med, LMA, Marseille, France\\
}
\date{Submitted to Acta Acustica, June 20, 2025\\
{Accepted July 28, 2025}}

\begin{document}
\maketitle
\begin{onecolabstract}
The Transfer Matrix Method is a practical approach for modeling plane wave propagation in one-dimensional waveguides.
Its simplicity makes it especially attractive for accounting for viscothermal losses, enabling realistic simulations of complex waveguides such as wind instruments.
Another strength of this method lies in its fully analytical formulation of wave propagation.
Modal parameters naturally arise as by-products of the model, obtained by numerically solving analytical expressions.
In this work, the analytical potential of the method is extended by deriving the sensitivity of modal parameters to changes in the geometry of the resonator.
These analytical gradients are applied in the context of wind instrument design.
A simplified model of a soprano saxophone is used to investigate how octave harmonicity can be optimized through small geometric adjustments.
The proposed approach enables predictive adjustments of geometry and offers valuable insight for both sound synthesis and instrument making.\\

\end{onecolabstract}
  \vspace{10pt}

\section{Introduction}\label{sec:introduction}

Tuning a wind instrument is a challenging task. 
A geometric modification that improves the intonation of one note may degrade another. 
Instrument makers rely on decades of empirical knowledge, trial and error, and fine craftsmanship. 
Acoustic improvements often come with ergonomic constraints. 
For instance, adding a tonehole may require extra keywork, which increases mechanical complexity, adds weight, and increases the risk of leakage.

To support the design process, researchers have developed analytical tools to quantify the acoustic impact of geometric modifications. 
A common approach models local discontinuities in the bore (such as toneholes or cross-sectional changes) as effective length corrections. 
The pioneering work of Nederveen (1969)~\cite{nederveen1969acoustical} provides analytical expressions for the length correction associated with various geometries, in the lossless case and for the first resonance frequency. 
These results were later consolidated and extended by Debut \textit{et al.}\ (2004)~\cite{debut_analysis_2005}, and applied to optimize the tuning of the second register of the clarinet. 

These tools are valuable for researchers. 
However, they may become inaccurate when more elaborate models accounting for viscothermal losses are introduced. 
They are also not always convenient for numerical implementation, particularly for non-specialists.

Length correction approaches typically focus on resonance frequencies. 
In many applications, however, it is essential to consider the full set of modal parameters, including damping and modal amplitudes. 
This is particularly relevant in the context of sound synthesis~\cite{guillemain2005real, taillard2018modal}, or for analyzing wind instruments as nonlinear dynamical systems~\cite{karkar2012oscillation, colinot2021multistability, terrien2013flute, doc2014minimal, freour2022parameter}.

The Transfer Matrix Method (TMM) provides a robust and practical alternative. 
It models one-dimensional waveguides by multiplying elementary sections in the frequency domain. 
Viscothermal and radiation losses can be easily included. 
The input impedance computed using TMM is expressed as a ratio of analytic functions. 
Its poles and residues approximate the modal parameters of the resonator. 
This approach is detailed in Chapter 2.2 of Debut (2004) \cite{debut2004deux} and applied in Taillard \textit{et al.} (2018)~\cite{taillard2018modal} and by the authors \cite{szwarcberg2023amplitude, szwarcberg2024second}.

Sensitivity analysis complements these methods by quantifying how small geometric changes affect the acoustic response. 
Nederveen~\cite{nederveen1969acoustical} first derived the sensitivity of resonance frequencies in the lossless case. 
Fachinetti \textit{et al.}\ (2003) \cite{facchinetti2003numerical} compute sensitivities of eigenfrequencies using a finite element model of the clarinet. 
In the context of bore reconstruction, Ernoult \textit{et al.}\ (2021) \cite{ernoult2021full} computed the sensitivity of impedance-based metrics with respect to geometric parameters, using finite element models.

In this work, we compute analytically the sensitivity of modal parameters for resonators modelled by the TMM. 
The formulation allows analytical expansions to arbitrary order, improving the accuracy of predictions for larger parameter variations. 
We provide a computational framework that supports practical implementation. 
The method is applied to the optimization of the second register intonation of a simplified soprano saxophone. 
We also demonstrate its application in time-domain sound synthesis when geometric parameters evolve over time.

\section{General method}
\subsection{Transfer {Matrix} modeling and modal decomposition}

In the frequency domain denoted by the Laplace variable $s$, the propagation of acoustic waves through a waveguide can be described using transfer matrices.  
For each element $i$, the transfer matrix $M_i$ relates the acoustic pressure and volume velocity at the upstream and downstream ends:

$$ 
\begin{pmatrix}
P_i \\ U_i
\end{pmatrix}
=
\underbrace{
\begin{pmatrix}
A & B \\ C & D
\end{pmatrix}
}_{M_i}
\begin{pmatrix}
P_{i+1} \\ U_{i+1}
\end{pmatrix}.
$$

The coefficients of $M_i$ depend on the geometry between positions $i$ and $i+1$.  
Transfer matrix models exist for cylindrical, conical, and flared sections, as well as for discontinuities such as side holes and diameter changes.

For each geometry, different models are available to approximate viscothermal losses with various degrees of accuracy.  
These losses are included through the complex wavenumber $\Gamma$.  
In the lossless case, assuming plane wave propagation, the wavenumber is simply:
$$
\Gamma(s) = \frac{s}{c_0},
$$
where $c_0$ is the speed of sound in the air.  
In more realistic models, $\Gamma$ depends on the waveguide geometry (e.g., bore diameter) and on the properties of the gas (temperature, humidity).

Propagation through the complete waveguide is described by the product of the transfer matrices of all $N$ elements:
$$
\begin{pmatrix}
P_1 \\ U_1
\end{pmatrix}
=
\underbrace{
\prod_{i=1}^{N} M_i
}_{M_\mathrm{tot}}
\begin{pmatrix}
P_{N+1} \\ U_{N+1}
\end{pmatrix}.
$$
The total transfer matrix $M_\mathrm{tot}$ has the form:
$$
M_\mathrm{tot} =
\begin{pmatrix}
A_\mathrm{tot} & B_\mathrm{tot} \\
C_\mathrm{tot} & D_\mathrm{tot}
\end{pmatrix}.
$$
The input impedance at the upstream end is then:
\begin{equation}\label{eq:zin}
Z_\mathrm{in} = \frac{A_\mathrm{tot} Z_R + B_\mathrm{tot}}{C_\mathrm{tot} Z_R + D_\mathrm{tot}},
\end{equation}
where $Z_R = P_{N+1} / U_{N+1}$ is the radiation impedance at the output of the waveguide.
This impedance can be expressed as a ratio of two analytic functions:
$$
Z_\mathrm{in} = \frac{\mathcal{N}(s)}{\mathcal{D}(s)}.
$$
The complex poles $s_n$ of the input impedance are the roots of the denominator:
\begin{equation}\label{eq:sn}
\mathcal{D}(s_n) = 0.
\end{equation}
In the absence of losses and for simple geometries, Eq.\ \eqref{eq:sn} can be solved analytically.
For more realistic models, it is solved numerically from appropriate initial conditions.

Each complex pole $s_n$ is related to a resonance frequency $f_n$, with:
$$
 2 \pi f_n = \Im(s_n).
$$
The residue $C_n$ at each pole is defined as the first-order coefficient in the Laurent expansion of $Z_\mathrm{in}$ around $s_n$:
\begin{equation}\label{eq:Cn}
C_n = \frac{\mathcal{N}(s_n)}{\mathcal{D}'(s_n)},
\end{equation}
where $\mathcal{D}'$ is the derivative of $\mathcal{D}$ with respect to $s$.  
Simple poles are assumed (i.e., $\mathcal{D}'(s_n) \neq 0$).
Note that a double pole in the input impedance means that two resonant modes coincide in frequency. 
Finally, the input impedance can be approximated by a modal decomposition involving $N_m$ modes:
\begin{equation}\label{eq:zmod}
Z_\mathrm{mod}(s) = \sum_{n=1}^{N_m} \left( \frac{C_n}{s - s_n} + \frac{\mathrm{conj}({C_n})}{s - \mathrm{conj}({s_n})} \right).
\end{equation}
Note, however, that the previous expression does not perfectly approximate the input impedance, even for $N_m \to \infty$. 
This is partly because $|Z_\mathrm{in}|$ does not vanish as $|s| \to \infty$, and also due to the presence of 
non-integer powers of $s$ in the complex wavenumber $\Gamma$, which can be written with fractional derivatives in the time domain.
The methods and conditions for computing the exact modal decomposition of the input impedance are detailed in Monteghetti (2018, Chap. 2.1) \cite{monteghetti2018analysis}.
An alternative approach to reduce the error caused by modal truncation involves adding a high-frequency corrective mode, as proposed by Guillemain and Silva (2010) \cite{guillemain2010utilisation}.

The extraction of complex poles $s_n$ and residues $C_n$ is essential in the physical modeling of wind instruments. 
In a wind instrument model written as a Cauchy problem, the resonator response can be written through a system of $N_m$ first-order ODEs. 
This representation is efficient for real-time synthesis or numerical continuation.

\subsection{Sensitivity of the poles to a parameter variation}
In wind {instrument} making, it can be difficult to assess intuitively how the resonance frequencies would shift when a parameter is slightly modified.
The analytical formalism of Transfer Matrices enables to answer this question elegantly.

Let us consider a waveguide made of $N$ different elements. 
The total TM of the waveguide $M_\mathrm{tot}$ is parametrized with respect to all geometric dimensions (lengths, radii, tonehole diameters and height{s}...), and to all characteristics of the ga{s} (for instance, temperature or humidity).
These parameters are gathered in a vector $\mathbf{\Theta}$.
We now focus on one parameter $\theta$ of the model.
Function $\mathcal{D}$ is viewed as a function of two variables, $s$ and $\theta$. 
The poles $s_n$ are implicitly defined with respect to $\theta$ by the equation
\begin{equation}\label{eq:1}
\mathcal{D}(s_n(\theta), \theta)=0.
\end{equation}
To determine how the poles vary with respect to $ \theta $, Eq.~\eqref{eq:1} is differentiated using the chain rule:
\begin{equation}
\frac{\mathrm{d}}{\mathrm{d}\theta} \mathcal{D}(s_n(\theta), \theta)
= \frac{\partial \mathcal{D}}{\partial s}(s_n, \theta) \frac{\mathrm{d}s_n}{\mathrm{d}\theta}
+ \frac{\partial \mathcal{D}}{\partial \theta}(s_n, \theta) = 0.
\end{equation}
Solving for $ {\mathrm{d}s_n}/{\mathrm{d}\theta} $ yields:
\begin{equation}
\frac{\mathrm{d}s_n}{\mathrm{d}\theta}
= - \frac{\partial \mathcal{D}/\partial \theta\, (s_n, \theta)}{\partial \mathcal{D}/\partial s\, (s_n, \theta)}.
\end{equation}
It is assumed that $ \mathcal{D} $ is differentiable with respect to both $s$ and $\theta$, and that $ [{\partial \mathcal{D}}/{\partial s}](s_n, \theta) \neq 0$, so that the implicit function theorem ensures the existence of a differentiable function $s_n(\theta) $ in a neighborhood of $\theta$.

\subsection{Sensitivity of the residues to a parameter variation}

We now study how $C_n$ varies with respect to a parameter $\theta$. 
Equation~\eqref{eq:Cn} becomes:
\begin{equation*}
C_n(s_n(\theta), \theta) = \frac{\mathcal{N}(s_n(\theta), \theta)}{\mathcal{D}'(s_n(\theta), \theta)}.
\end{equation*}
Differentiating with respect to $\theta$ gives:
\begin{equation}
\frac{\mathrm{d}C_n}{\mathrm{d} \theta} = 
\frac{\partial}{\partial s} \left[ \frac{\mathcal{N}}{\mathcal{D}'} \right](s_n, \theta) \cdot \frac{\mathrm{d}s_n}{\mathrm{d} \theta} 
+ \frac{\partial}{\partial \theta} \left[ \frac{\mathcal{N}}{\mathcal{D}'} \right](s_n, \theta).
\end{equation}
We assume that $\mathcal{N} \in \mathcal{C}^1$ and $\mathcal{D} \in \mathcal{C}^2$ with respect to $s$ and $\theta$.

\subsection{Practical implementation details}

Computing the sensitivities of poles and residues with respect to a parameter requires the evaluation of many derivatives.  
For a resonator with complex geometry and accurate loss models, deriving these expressions by hand is both tedious and difficult to maintain when the model evolves.
To address this, symbolic computation is used via Matlab's Symbolic Toolbox. The model is entirely defined using symbolic variables, allowing the sensitivity functions to be computed automatically.
However, symbolic expressions in Matlab are inefficient to evaluate numerically. To overcome this, the built-in \texttt{matlabFunction} command is used to generate optimized Matlab function files from the symbolic expressions. This step significantly improves performance and enables efficient use of the sensitivity functions within numerical loops.
A minimal working example can be downloaded from {Ref.\ \cite{szwarcberg2025minimal}}.

\section{Application: tuning of the second register of a soprano saxophone}

\begin{figure}[H]
	\centering
	\includegraphics[width=.9\textwidth]{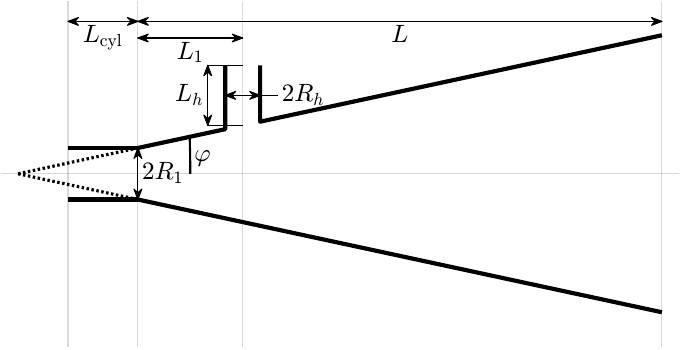}
	\caption{Scheme of the simplified saxophone.}
	\label{figsax:1}
\end{figure}

This section applies the sensitivity analysis of modal parameters to a practical acoustical problem: tuning the second register of a soprano saxophone.
We consider a simplified saxophone geometry, shown in Figure \ref{figsax:1}.
It is composed of three main elements:
\begin{itemize}
	\item a truncated cone of input radius $R_1$, half-angle $\varphi$ and total length $L$,
	\item a register hole, with radius $R_h$ and chimney height $L_h$, placed at a distance $L_1$ from the input radius of the truncated cone,
	\item a cylindrical mouthpiece, with radius $R_1$ and length $L_\mathrm{cyl}$.
\end{itemize}
The length $L_2=L-L_1$ can vary, allowing the instrument to play different notes. 
The length $L_\mathrm{cyl}$ is fixed so that the cylindrical mouthpiece occupies the same volume as the missing volume of the truncated cone:
\begin{equation}\label{eq:Lcyl}
L_\mathrm{cyl}= \frac{R_1}{3 \tan \varphi}.
\end{equation}

When the register hole is closed, the player produces notes near the first resonance frequency. 
This is the first register, noted $R1^{(c)}$ (closed hole).  
When the register hole is open, notes close to the second resonance frequency are preferably produced. This is the second register, noted $R_2^{(o)}$ (open hole). 
In an idealized conical resonator, the second register note is an octave above the first.

Instrument makers aim to tune both registers as closely as possible. 
A key challenge is to balance them across the full playing range. 
This is quantified by the inharmonicity, defined as:
\begin{equation}\label{eq:h}
h(s_1^{(c)}, s_2^{(o)} )= 1200 \log_2 \left[ \frac{\Im (s_2^{(o)})}{2 \Im (s_1^{(c)})} \right].
\end{equation}
Here, $s_1^{(c)}$ and $s_2^{(o)}$ are the poles of the input impedance when the register hole is closed and open, respectively.  
The function $h$ measures the deviation from an octave (in cents). 
A value of zero indicates perfect tuning.
The sensitivity of $h$ to a variation of $\theta$ is computed:
\begin{equation}\label{eq:sensih}
\frac{\mathrm{d}h}{\mathrm{d}\theta} = \frac{1200}{\log 2} \left[ \frac{1}{\Im(s_2^{(o)})} \cdot \Im \left( \frac{\mathrm{d} s_2^{(o)}}{\mathrm{d} \theta} \right) - \frac{1}{\Im(s_1^{(c)})} \cdot \Im \left( \frac{\mathrm{d} s_1^{(c)}}{\mathrm{d} \theta} \right) \right]
\end{equation}

The inharmonicity function is computed for {12} note pairs, from {D$_4$/D$_5$} to C$\sharp_5$/C$\sharp_6$.  
All notes are written in B$\flat$ transposition, as in a soprano saxophone fingering chart. 
Each pair consists of a first-register note and its corresponding second-register note.
{The lowest notes of the first register (B$\flat_3$ to C$\sharp_4$) are not studied here, as their corresponding octaves (B$\flat_4$ to C$\sharp_5$) are typically played with different first-register fingerings.
Similarly, the highest notes of the second register (D$_6$ to F$\sharp_6$) are excluded because their lower octaves (D$_5$ to F$\sharp_5$) are played with different second-register fingerings.
}

Instruments of the saxophone family feature two different register holes.
The first one is a {small-diameter} hole located close to the mouthpiece.
It opens from the A$_4$/A$_5$ to the {F$\sharp_5$/F$\sharp_6$}. 
The second one is located downstream from the first one. 
It opens from the bottom B$\flat_3$/B$\flat_4$ to the G$\sharp_4$/G$\sharp_5$.
Thus, the two register holes are never opened together.
The main parameters of the saxophone and the register holes are listed in Table~\ref{tabsax:1}.
They were measured on a real soprano saxophone.

\begin{table}[H]
\centering
\caption{Initial parameters of the simplified soprano saxophone. 
The temperature is fixed at $T_0=20^\circ$C.
The two cases correspond to the two register holes of the saxophone.}
\label{tabsax:1}
\begin{adjustbox}{width=1.0\textwidth}
\begin{tabular}{lcccccc}
\hline
 & $\varphi$ ($^\circ$) & $R_1$ (mm) & $L_1$ (mm) & $R_h$ (mm) & $L_h$ (mm) & Effectivity (Notes)\\
Upstream hole & 1.74 & 4.6 & 40 & 0.85 & 4.0 & $[$A$_4;~${F}$\sharp_5]$\\
Downstream hole & 1.74 & 4.6 & 130 & 1.4 & 4.0 & $[$B$\flat_3;~ $G$\sharp_4]$\\
\hline
\end{tabular}
\end{adjustbox}
\end{table}

\subsection{Preliminary validation of the method}

The reliability of the modal decomposition is first evaluated by comparing the modal impedance $Z_\mathrm{mod}$ with its analytical equivalent $Z_\mathrm{in}$. 
We then verify that modal coefficients can be explicitly estimated under small variations of geometric parameters.

The resonator is initially studied with closed register holes. 
Its complete transfer matrix, denoted $M_\mathrm{tot}^{(c)}$, is expressed as the product of four elementary matrices:
\begin{equation}
M_\mathrm{tot}^{(c)} = M_\mathrm{cyl} M_{c1} M_h^{(c)} M_{c2},
\end{equation}
where $M_\mathrm{cyl}$ corresponds to the cylindrical mouthpiece, $M_{c1}$ to the conical segment of length $L_1$, $M_h^{(c)}$ to the transfer matrix of the closed side hole, and $M_{c2}$ to the conical segment of variable length $L_2$.

References for each transfer matrix model, including viscothermal losses and radiation effects, are given in the Appendix. 
The chosen models introduce minimal approximations compared to commonly used alternatives in wind instrument modeling.

\begin{figure}[H]
	\centering
	\includegraphics[width=\textwidth]{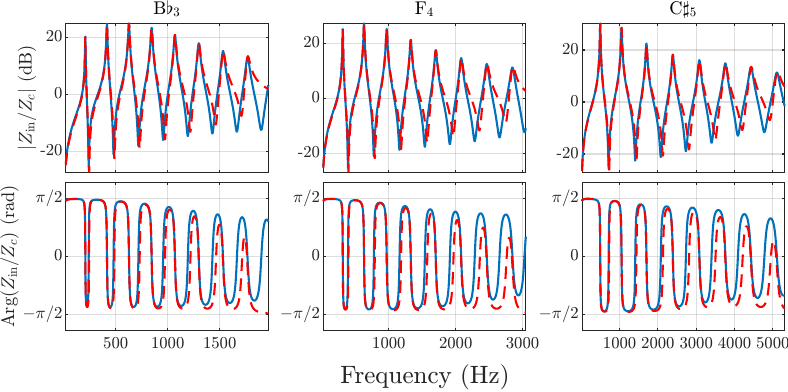}
	\caption{Modulus and phase of the input impedance.
	 Blue lines: direct computation from the TMM. 
	 Red {dashed} lines: modal decomposition with eight modes, obtained through Equations \eqref{eq:sn} and \eqref{eq:Cn}.
	Three notes of the first register are represented: B$\flat_3$ ($L=667$~mm), F$_4$ ($L=396$~mm) and C$\sharp_5$ ($L=197$~mm).}
	\label{figsax:zmod}
\end{figure}

Modal parameters are extracted in two steps. First, the complex poles $s_n$ are obtained by solving Eq.\ \eqref{eq:sn} numerically.
This first step is the only numerical step of the process.
Then, the residues $C_n$ are computed using Eq.\ \eqref{eq:Cn}. 
The modal impedance $Z_\mathrm{mod}$ [Eq.\ \eqref{eq:zmod}] is reconstructed and compared to the analytical impedance $Z_\mathrm{in}$ [Eq.\ \eqref{eq:zin}], as shown in Figure~\ref{figsax:zmod}. $N_m=8$ modes are included in $Z_\mathrm{mod}$.

Results are presented for three first-register notes: B$\flat_3$ ($L=667$~mm), F$_4$ ($L=396$~mm), and C$\sharp_5$ ($L=197$~mm). Resonance frequencies and amplitudes from $Z_\mathrm{mod}$ are compared to those from $Z_\mathrm{in}$.

For the first impedance peak, the frequency discrepancy remains below $0.1$~cent for all notes. 
Relative differences on the amplitude are below $0.03$\%. 
Discrepancies increase with the peak index. 
For the fourth peak, frequency and amplitude deviations remain below $0.6$~cent and $0.25$\%, respectively.
For the eighth peak, they are below $6.7$~cents and 11\%.
These discrepancies are reduced when more modes are accounted for.

Figure~\ref{figsax:zmod} also highlights discrepancies at antiresonances, especially for higher modes. 
These are significantly reduced by including more modes in the reconstruction.

The results confirm that the modal decomposition accurately fits the input impedance defined by the TMM. This approach provides a robust way to extract modal parameters, relying only on the derivative of the denominator of $Z_\mathrm{in}$ with respect to complex frequency.
This computation can be efficiently performed using symbolic algebra tools.

The sensitivity of the modal coefficients to a geometric parameter is now evaluated. The initial cone length is set to $L = 190$~mm. Modal frequencies $s_n$ and coefficients $C_n$ are first computed at this initial configuration. Their sensitivities with respect to the cone length, $\mathrm{d}s_n / \mathrm{d}L$ and $\mathrm{d}C_n / \mathrm{d}L$, are then calculated.

The cone length is progressively increased in steps of $\Delta L = 1$~mm, up to a final value of $L = 750$~mm. At each step, $s_n$ and $C_n$ are updated using a first-order perturbation:
\begin{equation}
s_n(L+\Delta L) = s_n(L) + \Delta L \, \frac{\mathrm{d}s_n}{\mathrm{d}L}(L).
\end{equation}
The sensitivity functions are recomputed at each step.
 
\begin{figure}[H]
	\centering
	\includegraphics[width=\textwidth]{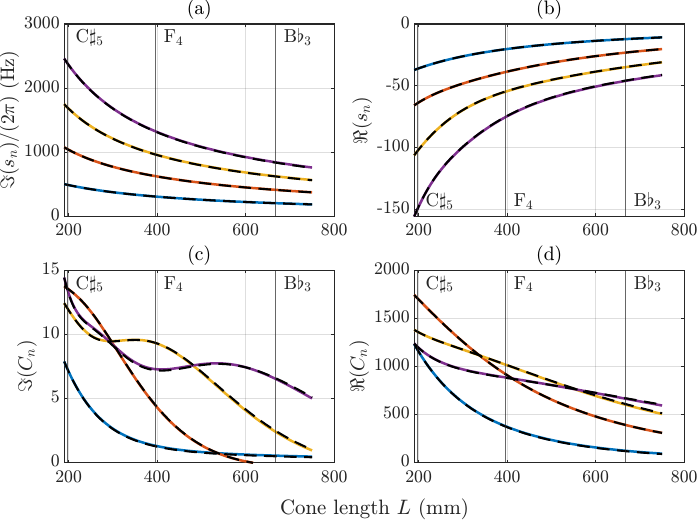}
	\caption{Variation of the modal coefficients $s_n$ and $C_n$, when the length of the cone $L$ increases.
	Blue, red, yellow and purple curves refer to modes 1 to 4 respectively. 
	Dashed lines show the explicit estimation of the modal coefficients through the sensitivity functions $\mathrm d s_n / \mathrm d L$ and $\mathrm d C_n / \mathrm d L$.
	}
	\label{figsax:reconstruction}
\end{figure}

Figure~\ref{figsax:reconstruction} shows the real and imaginary parts of the estimated modal coefficients (dashed lines) compared to their exact values (solid colored lines), obtained by solving Eq.\ \eqref{eq:sn} and applying Eq.\ \eqref{eq:Cn} for the first four modes ($N_m=4$).

Results indicate that the evolution of the poles $s_n$ is well captured by the first-order method. After 560 steps, the relative error on both $\Re(s_n)$ and $\Im(s_n)$ remains below 0.6\%. 

For the residues $C_n$, {close inspection of Figure \ref{figsax:reconstruction}(c,d) reveals that} the perturbation method leads to larger errors. For the first mode, the relative error reaches approximately 10\% in $\Im(C_n)$ and 4\% in $\Re(C_n)$ after 560 steps. When $C_n$ is recomputed from the estimated $s_n$ using Eq.\ \eqref{eq:Cn}, the error is reduced: less than 5\% in $\Im(C_n)$ and less than 2\% in $\Re(C_n)$.

Improved accuracy can be obtained with higher-order perturbation methods, or predictor-corrector methods.
Since derivatives of any order can be analytically computed, the refinement is straightforward in principle. 
However, the complexity of symbolic expressions increases, which may increase the computational cost.
This flexibility lets users select a method that matches their needs in terms of computational efficiency and convergence.

These sensitivity functions are especially valuable in time-domain simulations. When a geometric parameter varies over time, the modal equations can be dynamically updated using their sensitivities. This approach enables the modeling of systems with time-dependent geometry, such as the trombone slide, the \textit{glissotar} \cite{glissotar2024}, or opening side holes \cite{terroir2005simple}.

\subsection{Sensitivity curves for an intuitive optimization of second-register tuning}

\begin{figure}[H]
	\centering
	\includegraphics[width=\textwidth]{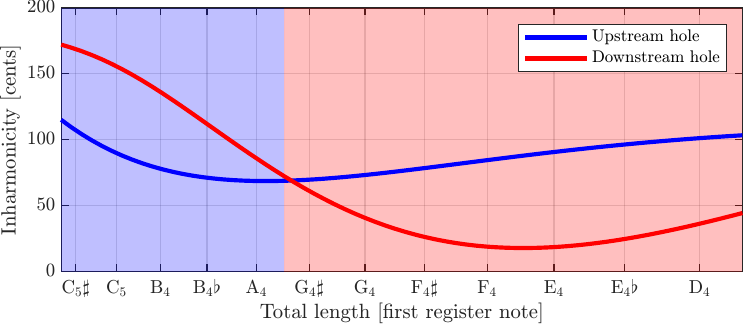}
	\caption{Inharmonicity $h$ between R2$^{(o)}$ and R1$^{(c)}$, for all notes between C$\sharp_5$ ($L=197~$mm) and {D$_4$} ($L=499$~mm).
	The blue and red curves show the evolution of $h$ when the upstream or downstream register hole is opened, respectively. 
	Colored areas indicate which hole is used.
	}
	\label{figsax:h}
\end{figure}

Figure~\ref{figsax:h} shows the inharmonicity $h$ for each register hole.  
All second-register notes are too high in pitch, particularly the highest ones.  
The minimum inharmonicity for the upstream hole is $+68.8$~cents, reached at A$_4$/A$_5$.  
For the downstream hole, it is $+18.6$~cents at F$_4$/F$_5$.

Before proposing strategies to reduce inharmonicity, the relationship between the register hole positions and the pressure node of the second mode is briefly examined.

\begin{figure}[H]
	\centering
	\includegraphics[width=.7\textwidth]{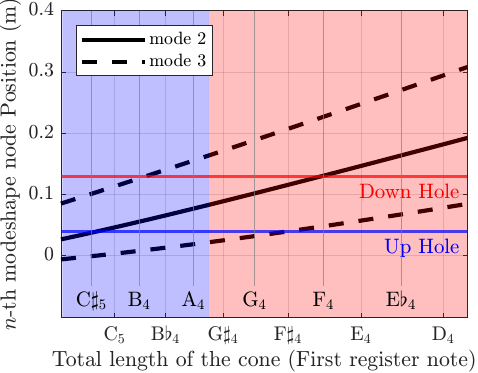}
	\caption{Location of pressure nodes for the second (solid black line) and third mode (dashed black lines) with closed holes.  
	Horizontal blue and red lines indicate the position of the upstream and downstream register holes, respectively.
	Colored areas indicate positions where each hole is opened.
	Details on the computation of mode shapes using the TMM can be found in the Appendix C of Szwarcberg \textit{et al.}\ (2024) \cite{szwarcberg2024second}.}
	\label{figsax:mode}
\end{figure}

Figure~\ref{figsax:mode} shows the position of the second mode shapes' pressure node for all first-register fingerings, in solid black line.  
For C$\sharp_5$, this node aligns with the upstream register hole.  
For F$_4$, it aligns with the downstream hole.  

This also explains why the register hole switch occurs between G$\sharp_4$ and A$_4$.  
For the G$_4$ fingering, the upper register hole is located near a pressure node of the third mode (dashed lines).
Opening the upper register hole for this note would favor the third register (D$_6$).
{For the A$_4$ fingering, and for higher notes, the upstream register hole is positioned closer to a pressure node of the second mode than the downstream hole.
As a result, it provides better tuning in the second register, even though both register holes are near a pressure node of the third mode.
When extending the curves to the lowest notes (C$\sharp_4$ to B$\flat_3$), we observe that opening the downstream register hole in these cases may favor the third register rather than the second.
}

\begin{figure}[H]
	\centering
	\includegraphics[height=4.2cm]{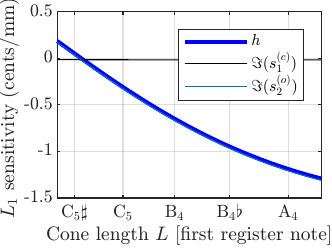}
	\hspace{0.8cm}
	\includegraphics[height=4.2cm]{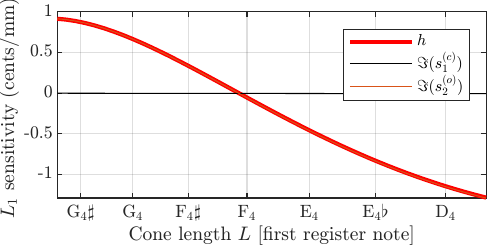}\\
	\includegraphics[height=4cm]{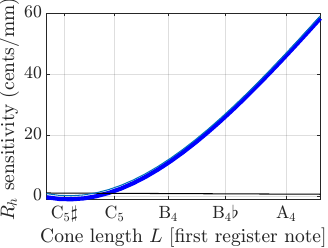}
	\hspace{1cm}
	\includegraphics[height=4cm]{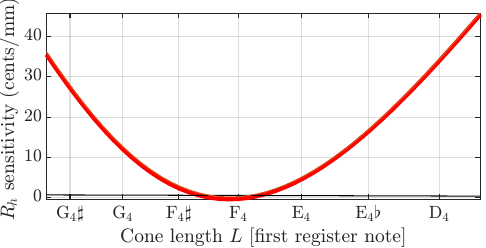}\\
	\includegraphics[height=4cm]{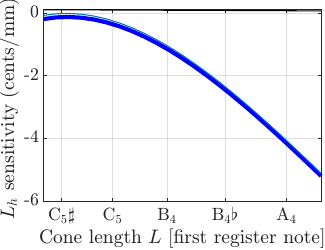}
	\hspace{1cm}
	\includegraphics[height=4cm]{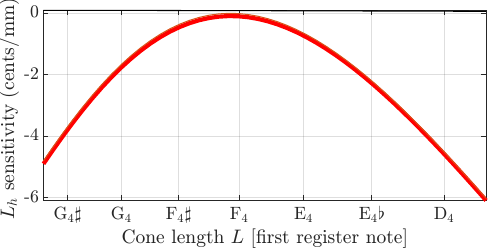}
	\caption{Sensitivity of the inharmonicity $h$ to small changes in $L_1$, $R_h$, and $L_h$ (in cents per mm).  
	Rows correspond to the parameter varied.  
	Left column: upstream hole. Right column: downstream hole.
	{Thin black ($y\approx 0$): sensitivity of first-register notes.
	Thin colored: second-register notes.
	Thick colored: difference between second- and first-register sensitivities.
	The thin and thick colored lines almost completely overlap, reflecting the fact that changes to a closed register hole have minimal impact on the resonator’s resonance frequencies.
	}
	}
	\label{figsax:sensi}
\end{figure}

These mode-shape observations help interpret the sensitivity curves shown in Figure~\ref{figsax:sensi}, which are computed using Eq.\ \eqref{eq:sensih}.  
They provide intuitive guidelines to reduce inharmonicity.
{In these curves, a negative sensitivity value means that an increase in the parameter leads to a decrease in harmonicity, while a positive value implies the opposite trend.
}

\begin{itemize}
\item \textbf{Position of the register hole ($L_1$).}  
The first row of Figure~\ref{figsax:sensi} shows the sensitivity of $h$ with respect to $L_1$.  
{For the upstream register hole (blue curve), increasing} $L_1$ reduces inharmonicity for notes lower or equal than C$_5$/C$_6$, and increases inharmonicity for the C$\sharp_5$/C$\sharp_6$.
Figure~\ref{figsax:mode} shows that the sensitivity crosses zero when the {upper} register hole coincides with the second mode shape's pressure node.
This effect is expected: moving the {register} hole closer to a pressure node improves harmonicity.  
{Likewise, in the case of the downstream hole (red curve), the register hole coincides with the pressure node of the second mode shape for note F$_4$.
Consequently, increasing $L_1$ leads to greater inharmonicity for notes above F$_4$, and reduced inharmonicity for those below.} 

\item \textbf{Radius of the register hole ($R_h$).}  
The second row shows that {increasing $R_h$ increases inharmonicity} across most notes, except when the hole coincides with a pressure node of the second mode shape.  
{Conversely, decreasing $R_h$ improves tuning.}
However, if $R_h$ is too small, the second register cannot be played.  
The authors observed \cite{szwarcberg2024second} that for $R_h \leq 0.5$~mm on a cylindrical resonator, opening the hole fails to overblow to the second register and stays on the first, R1$^{(o)}$.

\item \textbf{Chimney height ($L_h$).}  
As shown in the third row, increasing $L_h$ reduces inharmonicity, similarly to decreasing $R_h$.
\end{itemize}

\begin{figure}[H]
	\centering
	\includegraphics[width=.39\textwidth]{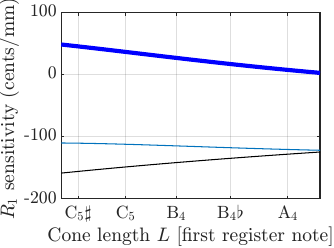}
	\includegraphics[width=.58\textwidth]{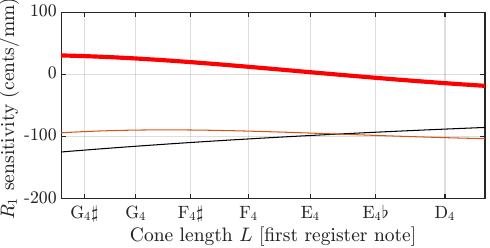}\\
	\includegraphics[width=.39\textwidth]{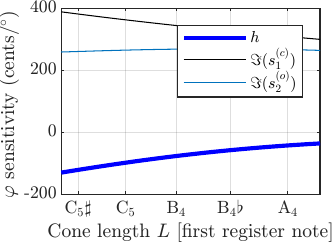}
	\includegraphics[width=.58\textwidth]{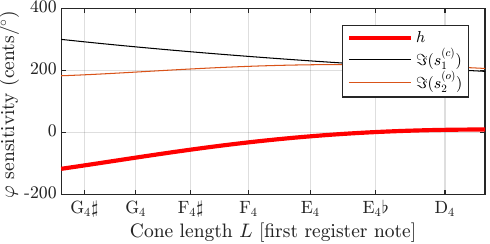}
	\caption{Sensitivity of inharmonicity $h$ to small changes in $R_1$ (first row) and $\varphi$ (second row).  
	Left: upstream hole. Right: downstream hole.  
	Thin black: sensitivity of first-register notes.  
	Thin colored: second-register notes.  
	Thick colored: difference between second- and first-register sensitivities.}
	\label{figsax:sensi2}
\end{figure}

Additional modifications, such as adjusting the input radius $R_1$ or the cone half-angle $\varphi$, are presented in Figure~\ref{figsax:sensi2}.  
These changes could improve tuning but would require a complete redesign of the instrument.

\begin{itemize}
\item \textbf{Input radius ($R_1$).}  
Increasing $R_1$ lowers first-register resonance frequencies (R1$^{(c)}$) due to an increased effective mouthpiece length $L_\mathrm{cyl}$ and larger radiation correction{: as $\varphi$ remains constant, the radius at the open end increases.}
The second register (R2$^{(o)}$) is differently affected.  
Overall, raising $R_1$ improves harmonicity of lower second-register notes (below {E$_4$/}E$_5$) but worsens it for higher ones.

\item \textbf{Cone half-angle ($\varphi$).}  
Increasing $\varphi$ raises the first-register resonance frequencies by reducing $L_\mathrm{cyl}$.  
It reduces inharmonicity for high notes in the second register but has little effect below E$_4$/E$_5$.
Note that modern saxophones have a less pronounced conicity than older models \cite{postmaSaxophoneAcoustics}.
\end{itemize}

\subsection{An example of sensitivity-informed optimization}

Figure~\ref{figsax:h} shows that the largest inharmonicity is found at the edges of the range of the second register.  
These regions should be prioritized for improvement.

According to the sensitivity curves in Figures~\ref{figsax:sensi} and~\ref{figsax:sensi2}, simple improvements can be obtained by reducing the register hole radius and increasing the chimney height.  
These modifications are particularly effective for the downstream register hole.

\begin{figure}[H]
	\centering
	\includegraphics[width=\textwidth]{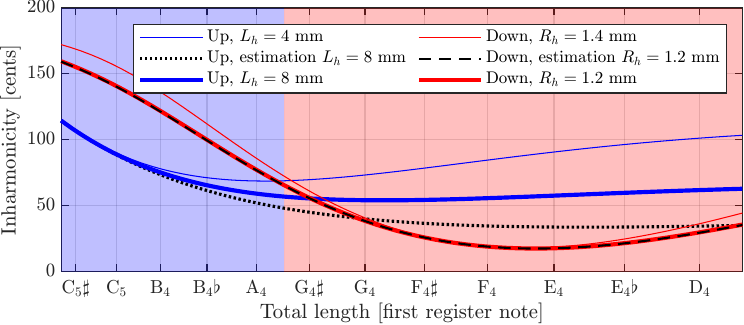}
	\caption{Incremental optimization of inharmonicity using sensitivity curves.  
	For the upstream register hole, $L_h$ is increased from 4~mm to 8~mm.  
	For the downstream hole, $R_h$ is reduced from 1.4~mm to 1.2~mm.
	{Thin colored lines represent the initial inharmonicity of the saxophone, prior to any parameter modifications.  
Black lines indicate the estimated correction obtained using first-order sensitivities [Eq.~\eqref{eq:sensih}].  
Thick colored lines correspond to the inharmonicity computed by directly solving Eq.~\eqref{eq:sn} with the modified parameters.
 }
	 }
	\label{figsax:optimh}
\end{figure}

A reduction of $R_h$ is tested for the downstream hole.  
From Figure~\ref{figsax:sensi}, { sensitivity functions estimate} that for {D$_4$/D$_5$}, decreasing $R_h$ by 0.1~mm lowers inharmonicity by {3.4}~cents.  
A total reduction of 0.2~mm is applied.  
{The expected correction is shown as a dashed black line in Figure~\ref{figsax:optimh}. 
It closely matches the direct computation of $h$ [Eq.~\eqref{eq:h}], obtained by solving Eq.~\eqref{eq:sn} using the TMM resonator model.
}

To improve the tuning of the highest second-register notes, the upstream hole’s chimney length is increased.  
According to Figure~\ref{figsax:sensi}, a 1~mm increase in $L_h$ leads to a 4-cent reduction for A$_4$/A$_5$.  
A 4~mm increase is tested.  
The predicted correction is shown as a dotted black line in Figure~\ref{figsax:optimh}.  
However, the actual inharmonicity (thick blue line) deviates from the estimate.  
This illustrates the limitation of the sensitivity approach: first-order approximations may become inaccurate for large parameter variations (here, an increase of 100\% from the initial value).
For more accurate predictions, higher-order corrections could be computed, as well as predictor-corrector methods.

Finally, the sensitivity curves show that {reducing inharmonicity for the highest notes (B$_4$/B$_5$ to C$\sharp_5$/C$\sharp_6$) can only be achieved by increasing $\varphi$ and decreasing $R_1$.}

Overall, sensitivity curves offer a powerful predictive tool for instrument makers, especially for guiding small geometric modifications.

\subsection{Discussion}

Sensitivity curves show that increasing the chimney height $L_h$ and reducing the radius $R_h$ can further improve tuning across most of the second register.  
However, these changes may compromise the ability of the instrument to switch registers.  
In particular, they increase the risk that the first-register regime remains stable even when the register hole is open.

Determining which oscillating regimes are stable for a given set of control parameters is not straightforward.  
This requires specific techniques such as continuation methods \cite{colinot2021multistability} or cartography of oscillating regimes via time-domain simulations \cite{colinot2025cartography, doc_oscillation_2015}.
However, Bouasse-Benade's prescription \cite{benade1959woodwind} can provide guidelines: an oscillation regime is favored when the resonance frequencies of the resonator are aligned with its harmonic series.  
In this context, when the register hole is opened, the emergence of the second register is favored when the inharmonicity between the first two impedance peaks is high.
Therefore, studying the sensitivity of the ratio $\Im(s_2^{(o)}) / \Im(s_1^{(o)})$ to a modification of a geometric parameter could be relevant.
The sensitivity to other modal parameters, such as the modal damping of the first mode $\xi_1$, with $$\xi_1 = |\Re(s_1)/s_1|, \quad \mathrm{and} \quad s_1=-\omega_1 \xi_1 + j\omega_1\sqrt{1-\xi_1^2},$$ could also be studied to determine the stability of the first register when the hole is opened.

Furthermore, localized nonlinear losses in the register hole should be considered when studying the playability of the second register \cite{szwarcberg2024second}.
If they are neglected, the model may produce a stable first register, even though it would be unstable in a real instrument.
Localized nonlinear losses can be modeled as a resistive impedance added to the shunt impedance of the hole \cite{dalmont_experimental_2002,debut_analysis_2005}.  
This resistance increases with the amplitude of the acoustic velocity inside the hole, $V$.
Note, however, that the sensitivity of $C_n$ and $s_n$ with respect to $V$ can be easily computed.

To conclude, sensitivity curves are useful to optimize the tuning of an instrument, but are not sufficient to quantify the stability of the oscillating regimes.

\section{Conclusion}
The modal parameters of any resonator defined by the Transfer Matrix Method can be obtained easily by {differentiating} the transfer matrices with respect to the Laplace variable.
The sensitivity of the modal coefficients with respect to geometric parameters of the resonator can then be determined analytically. 
In this article, first-order sensitivities are computed, although analytical developments can be extended to any order, allowing for higher-order corrections when the parameter variation is not small. 

The first interest of these functions is to recompute directly the modal coefficients when a parameter is slightly changed. 
This can be useful for physical model-based sound synthesis in which geometric parameters vary with respect to time (for instance: slide of a trombone, glissotar modeling, side hole opening).
Secondly, the sensitivity with respect to geometric parameters is a valuable tool for wind instrument makers.  
It helps to predict how a small change in geometry affects a performance-relevant quantity, such as inharmonicity between registers.

The method is applied here to analyze and improve the tuning between the first and second registers of a soprano saxophone.  
Sensitivity curves guide intuitive and quantitative design choices to reduce inharmonicity across the instrument's range.  
For instance, they highlight the influence of global geometric parameters, such as input radius $R_1$ or cone half-angle $\varphi$, although modifying these would require a significant redesign of the instrument.
  
However, sensitivity curves alone are not sufficient to predict how the stability of an oscillating regime is affected by a geometric change.  
A promising direction would be to integrate these sensitivity functions into a saxophone system written under the Harmonic Balance Method  \cite{gilbert1989calculation}.  
This would enable the computation of the stability of the oscillating regimes with respect to geometric modifications, and thus provide a complete framework for informed instrument design.

\section*{Declaration of competing interest}
The authors declare that they have no known competing financial interests or personal relationships that could have appeared to influence the work reported in this paper.

\section*{Acknowledgments}
This study has been supported by the French ANR LabCom LIAMFI (ANR-16-LCV2-007-01). 
The authors warmly thank M.\ Nouveau and K.\ Manachinskaia for the measurements of the soprano saxophone.

\section*{Data availability}
{A minimal working example can be downloaded from Zenodo: \url{https://doi.org/10.5281/zenodo.16150126}. \cite{szwarcberg2025minimal}.}
The data that support the findings of the present study is available from the corresponding author upon reasonable request.

\appendix
\section*{Appendix}\label{app}
\subsection*{Model}
This section briefly presents the components of the simplified saxophone described in Figure \ref{figsax:1}.
The complete model is made of a transfer matrix $M_\mathrm{tot}$, so that
$$ M_\mathrm{tot} = M_\mathrm{cyl} M_{c1} M_h M_{c2}.$$
\begin{itemize}
\item The expression of $M_\mathrm{cyl}$ is given in Chap.\ 4.5 of Chaigne and Kergomard (2016)\cite{bible2016}.
\item The expression of $M_\mathrm{hole}$  can be found in Lefevbre and Scavone (2012) \cite{lefebvre2012characterization}.
\item The expression of the transfer matrix of a cone can be found in Tournemenne and Chabassier (2019) \cite{tournemenne2019comparison}.
\item A complex wavenumber $\Gamma=\sqrt{Z_v Y_t}$ accounts for viscous ($Z_v$) and thermal $(Y_t)$ dissipation.
It is recalled in Tournemenne and Chabassier (2019) \cite{tournemenne2019comparison}.
The propagation medium is made of air at a temperature $T_0=20^\circ$C.
For the cylindrical part, $R^\odot=R_1$ for the computation of $\Gamma$.
\item Radiation is taken into account at the open end of the tube and in the register hole, when it is opened. 
The expression of the radiation impedance $Z_R$ can be found in Equation (9) of Silva \textit{et al.}\ (2009) \cite{silva_approximation_2009}.
We assume that the tonehole is infinitely flanged, and the open end is unflanged.
\end{itemize}

\printbibliography

\end{document}